%% file: paper.tex
\documentclass[preprint,numbers,10pt]{sigplanconf}

\usepackage{balance}
\usepackage{amsmath}
\usepackage{amsthm}
\usepackage{verbatim}
\usepackage{amsfonts}
\usepackage{amssymb}
\usepackage{graphics}
\usepackage{listings}
\usepackage{xspace}
\usepackage{ottlayout}
\usepackage{xcolor}
\usepackage{rotating}
\usepackage{alltt}
\usepackage{wrapfig}
\usepackage{hhline}
\usepackage{prettyref}
\usepackage{amscd}
\usepackage{multirow}
\usepackage{fancyvrb}
\usepackage{subfloat}
\usepackage{url}
\usepackage[override]{cmtt}
\usepackage{graphicx}
\usepackage{caption}
\usepackage{framed}
\usepackage{algorithm}
\usepackage[noend]{algpseudocode}
\usepackage{adjustbox}
\usepackage{mdframed}
\usepackage{minibox}
\usepackage{tikz}
\usepackage{footnote}
\usepackage{amsmath,amssymb}

\newrefformat{sec}{Section~\ref{#1}}
\newrefformat{subsec}{Section~\ref{#1}}
\newrefformat{fig}{Figure~\ref{#1}}
\newrefformat{tab}{Table~\ref{#1}}

\newcommand{\eg}{\emph{e.g.}\xspace}
\newcommand{\ie}{\emph{i.e.}\xspace}

\newcommand{\infer}{\textsc{Alive-Infer}\xspace}
\newcommand{\alive}{Alive\xspace}
\newcommand{\alivenj}{Alive-NJ\xspace}

\newcommand{\bnfdef}{\mathrel{\!\colon\!\!\colon\!\mathord{=}}}
\newcommand{\bnfalt}{\;\mathrel{\mid}\;}
\newcommand{\ottnt}[1]{\mathit{#1}}
\newcommand{\ottmv}[1]{\mathit{#1}}
\newcommand{\ottkw}[1]{\mathbf{#1}}

\newcommand{\runtimevars}{\mathcal{R}}
\newcommand{\compiletimevars}{\mathcal{C}}
\newcommand{\vin}{\runtimevars, \compiletimevars}
\newcommand{\val}{\iota}
\newcommand{\srcval}{\val_s}
\newcommand{\tgtval}{\val_t}

\newcommand{\welldef}{\delta}
\newcommand{\srcdef}{\welldef_s}
\newcommand{\tgtdef}{\welldef_t}

\newcommand{\safe}{\sigma}

\newcommand{\tgtsafe}{\safe_t}
\newcommand{\presafe}{\safe_\pre}

\newcommand{\subst}[3]{#1[#2/#3]}

\newcommand{\pre}{\phi}

\newcommand{\accepts}[2]{\Call{accepts}{#1,#2}}
\DeclareMathOperator*{\argmax}{argmax}

\definecolor{lightgray}{rgb}{0.9,0.9,0.9}

\usepackage{hyperref}
\begin{document}

\title{Precondition Inference for Peephole Optimizations in LLVM}
\authorinfo{David Menendez}{Rutgers University}{davemm@cs.rutgers.edu}
\authorinfo{Santosh Nagarakatte}{Rutgers University}{santosh.nagarakatte@cs.rutgers.edu}

\maketitle

\begin{abstract}
Peephole optimizations are a common source of compiler bugs. Compiler
developers typically transform an incorrect peephole optimization into
a valid one by strengthening the precondition. This process is
challenging and tedious. This paper proposes \infer, a data-driven
approach that infers preconditions for peephole optimizations
expressed in Alive. \infer generates positive and negative examples
for an optimization, enumerates predicates on-demand, and learns a set
of predicates that separate the positive and negative examples. \infer
repeats this process until it finds a precondition that ensures the
validity of the optimization. \infer reports both a weakest
precondition and a set of succinct partial preconditions to the
developer.
Our prototype generates preconditions that are weaker than LLVM's
preconditions for 73 optimizations in the Alive suite. We also
demonstrate the applicability of this technique to generalize 54
optimization patterns generated by Souper, an LLVM~IR--based
superoptimizer.
\end{abstract}

\input{sec.intro.tex}
\input{sec.background.tex}

\input{sec.approach.tex}
\input{sec.generalize.tex}
\input{sec.evaluation.tex}
\input{sec.related.tex}

\input{sec.conclusion.tex}
\section*{Acknowledgments}
We thank Aarti Gupta, Adarsh Yoga, Jay Lim, and the PLDI reviewers for
their feedback on drafts of this paper.  We thank John Regehr for his
blog posts on Souper. This paper is based on work supported in part by
NSF CAREER Award CCF--1453086, a sub-contract of NSF Award
CNS--1116682, a NSF Award CNS--1441724, a Google Faculty Award, and
gifts from Intel Corporation.
\balance
\bibliographystyle{myabbrvnat}
\bibliography{verifiedllvm}

\end{document}

%% file: sec.intro.tex
\section{Introduction}
\label{sec:intro}
LLVM is a widely used compiler, both in industry and academia. To attain
the best possible performance, LLVM performs a large number of
semantics-preserving optimizations. Among these are peephole optimizations,
which perform local rewriting of code with a primary focus on algebraic
simplifications. They also clean up and canonicalize code, which can
enable other optimizations. In LLVM, peephole optimizations find code
fragments in an input program that match a pattern, and replace them with
an equivalent set of instructions.
They are also a persistent source of LLVM
bugs~\cite{yang:csmith:pldi:2011,le:orion:pldi:2014,lopes:alive:pldi:2015}.

We have addressed the problem of peephole optimization bugs with
Alive, a domain-specific language for specifying and verifying
peephole optimizations in LLVM~\cite{lopes:alive:pldi:2015}.
The Alive language is similar to the LLVM intermediate representation
(IR).
An Alive optimization has a source pattern and a target pattern, with
an optional precondition~(see Section~\ref{sec:background}). The Alive
interpreter checks the correctness of an optimization using
satisfiability modulo theories (SMT) solvers. Alive has discovered
numerous bugs and is currently used by LLVM
developers~\cite{lopes:alive:pldi:2015,AppleBug, Patch1,
  AliveFPPatch}.

Alive prevents the inclusion of wrong optimizations in the LLVM
compiler. It also provides counterexamples for wrong optimizations.
The developer must exclude all inputs that make the optimization
invalid. Developers typically accomplish this by strengthening the
precondition of the optimization.
However, developing an appropriate precondition when presented with a
Alive counterexample can be tedious.  To illustrate, let us consider
the following peephole optimization~(presented in Alive syntax), which
was submitted as a code patch for the LLVM
compiler~\cite{AppleBug}:
\begin{verbatim}
Pre: isPowerOf2(C1 ^ C2)
%x = add %A, C1
%i = icmp ult %x, C3
%y = add %A, C2
%j = icmp ult %y, C3
%r = or %i, %j
  =>
%and = and %A, ~(C1 ^ C2)
%lhs = add %and, umax(C1, C2)
%r = icmp ult %lhs, C3
\end{verbatim}
The patch was rejected because Alive found it to be invalid and
provided a counterexample. After multiple revisions, the developer
found a precondition that made the optimization valid:
\begin{small}
\begin{verbatim}
C1 u> C3 && C2 u> C3 && abs(C1-C2) u> C3 &&
isPowerOf2(C1 ^ C2) && isPowerOf2(-C1 ^ -C2) &&
(-C1 ^ -C2) == ((C3-C1) ^ (C3-C2)) 
\end{verbatim}
\end{small}

The precondition of an optimization is a collection of predicates
involving symbolic constants, constant expressions, and constant
functions~(see Figure~\ref{fig:syntax}). It determines when the
optimization can be applied to an input program.
A strong precondition prevents the application of the optimization
for some programs where it would be valid.
For example, the precondition in the LLVM patch above rules out many
valid valuations for the symbolic constants.  A weaker precondition
that has too many predicates~(not succinct) can increase compilation
time, because the precondition has to be evaluated for every potential
application site. The optimization developer considers
trade-offs between the strength of the precondition and its
succinctness. Hence, identifying preconditions for these
optimizations is challenging.

This paper proposes \infer, a data-driven approach for identifying
appropriate preconditions for LLVM peephole optimizations expressed in
Alive.
\infer is inspired by PIE (Precondition Inference Engine)~\cite{padhi:pie:pldi:2016} and
other data-driven approaches~\cite{garg:ice:cav:2014,
  garg:ice-dt:popl:2016, sharma:guess-and-check:esop:2013,
  martin:commutative-spec:cav:2015, sankaranarayan:dt:issta:2008} to
generate preconditions and loop invariants for general purpose
programs.
We design new techniques and adapt the PIE approach to
address the following challenges in the Alive context.  First, we must
generate examples (data) in a static setting to use a data-driven
approach.
Second, we must address compile-time undefined behavior in the
constant expression language.
We want to reason about potentially unsafe predicates without
the risk of crashing LLVM at compile-time.
Third, we must handle type polymorphism in Alive while generating
examples and enumerating predicates.

\infer addresses these challenges and
proposes an end-to-end solution for generating
preconditions that developers can use.
We divide the task of inferring preconditions for an optimization into
three subtasks: (1) example generation, (2) predicate enumeration and
learning, and (3) Boolean formula learning.
\infer's example generator creates positive and negative examples for
an optimization. We propose that an example in this setting should
provide types and valuations for the symbolic constants.
\infer must consider types while generating examples because Alive
optimizations are parametric types.
It must ensure that no positive example causes compile-time undefined
behavior.
\infer generates examples through random selection and by querying an
SMT solver.
It classifies an example as positive if the refinement check for the
optimization is valid on substituting the symbolic constants with
concrete values, which is checked using an SMT solver. The refinement
check has a for-all quantification for the input variables.
To ensure sufficient number of positive and negative examples, \infer
supplements the randomly-chosen examples with examples obtained using
an SMT solver~(see Section~\ref{sec.example-generation}).

Next, \infer enumerates and learns a set of predicates to accept all
positive examples and reject all negative examples.
As with PIE~\cite{padhi:pie:pldi:2016}, 
new predicates are learned on-demand by enumerating predicates and
evaluating them on a small sample of examples.%, similar to
The enumerator lazily produces polymorphic, type-correct predicates.
\infer must consider whether a predicate can be evaluated safely.
For a given example, a predicate may be true, false, or unsafe.
\infer learns predicates which separate the positive and negative
examples in the sample. These narrow the search for future
predicates, and may be used in the
precondition~(see Section~\ref{sec.on-demand-learning}).

Once \infer learns sufficiently-many predicates, it uses one of
two Boolean formula learners to assemble a precondition.
The full Boolean learner produces preconditions which accept all
positive examples and reject all negative examples, but may be
large and complex. In contrast,
the partial Boolean learner produces succinct preconditions which
may not accept all positive examples. \infer reports multiple
partial preconditions to the developer as they are obtained,
and terminates once it produces a full precondition that is
proven to accept all positive examples, or \emph{weakest}.
This provides developers a choice between applicability and succinctness.
The Boolean learner must ensure compile-time safety of the learned
precondition. The presence of unsafe predicates introduces new
challenges in formula learning as the negation of an unsafe
predicate is still unsafe~(see Section~\ref{sec.boollearner}).
A partial precondition accepts a subset of the positive examples
while rejecting all negative examples.  The Boolean learner attempts
to maximize the number of positive examples accepted while generating
partial preconditions.
In contrast, a weakest precondition accepts all positive examples
and rejects all negative examples.
\infer checks the validity of both partial and weakest
preconditions, and whether a proposed weakest precondition rejects any
positive examples.

We built the \infer prototype for generating preconditions by
extending the publicly-available \alivenj toolkit~\cite{alivenj-git}.
We evaluated it using the Alive suite of optimizations. Out of the 417
optimizations in the Alive suite, there are 174 optimizations that
have a precondition. The \infer prototype generates the weakest
precondition for 133 of them within 1000 seconds. It generates either
a partial or the weakest precondition for 164 out of the 174
optimizations.  \infer generates a weaker precondition than the
precondition in the Alive suite for 73 optimizations.  We have also
used \infer to generalize concrete optimization patterns generated by
Souper~\cite{souper}, an LLVM~IR--based superoptimizer~(see
Section~\ref{sec:generalize}). We generalized a total of 71
optimizations that are expressible in Alive. \infer is able to
generate preconditions for 54 of them.

%% file: sec.background.tex
\section{Background on Alive}
\label{sec:background}

Alive~\cite{lopes:alive:pldi:2015} is a domain-specific language for
specifying and verifying peephole optimizations in LLVM. The Alive
interpreter checks the correctness of an Alive optimization by
encoding it as constraints, which allows automated reasoning with SMT
solvers. The interpreter also generates C++ code when the optimization
is correct. To encourage adoption by LLVM developers, the Alive
language is similar to the LLVM intermediate representation. Alive has
found numerous bugs in the LLVM
compiler~\cite{lopes:alive:pldi:2015,AppleBug}.  LLVM developers are
actively using Alive to check the correctness of new optimizations
(patches) submitted to LLVM. Alive based tools have prevented many
bugs in patches committed to LLVM~\cite{AppleBug, Patch1,
  AliveFPPatch}.
Although C++ code generation is not actively used,
there are plans for replacing InstCombine with Alive-generated C++
code.  Alive has also been extended to reason about the correctness of
floating point optimizations~(\eg,
Alive-FP~\cite{menendez:alive-fp:sas:2016} and
LifeJacket~\cite{lifejacket}).
We describe the language and the verification process next.

\paragraph{The Alive language.} An Alive optimization has the form
\texttt{source $\Rightarrow$ target}, with an optional
precondition. The source and target describe directed, acyclic
graphs~(DAGs) of values. Semantically, an Alive optimization replaces
the DAG in the source with the DAG in the target.
The interior nodes correspond to Alive~(LLVM
IR) instructions, with incoming edges representing their arguments.
Leaf nodes are input variables. They represent arbitrary LLVM values,
such as results from other instructions, symbolic constants, constant
expressions, and function parameters.
We will write $\compiletimevars$ for the set of input variables whose
values are known while compiling a concrete input program. Types for
the variables, values for symbolic constants and constant expressions
are available during compilation.  We use $\runtimevars$ to represent
remaining input variables that are not known at compile time.

An example Alive optimization is shown in
Figure~\ref{fig:illustration}(a) and its DAG representation is shown
in Figure~\ref{fig:illustration}(b). There are three input variables:
\texttt{\%X}, \texttt{C1}, and \texttt{C2}, where \texttt{\%X} is a 
run-time input variable, \texttt{C1} and \texttt{C2}
are symbolic constants, and \texttt{C1 /u C2} is a constant
expression.

\input{precondition.tex}
\paragraph{Preconditions.}
A precondition for a peephole optimization in LLVM is checked during
compilation of an input program before applying the
optimization. Hence, preconditions for these optimizations primarily
deal with values that can be determined during compilation: types,
symbolic constants, and constant expressions. Figure~\ref{fig:syntax}
provides the abstract syntax of preconditions for LLVM peephole
optimizations. A precondition is a conjunction or disjunction of
various predicates. A predicate is either a predicate function or a
binary comparison operation involving constant expressions. Constant
expressions can be symbolic constants, constant functions, and binary
operations of constant expressions.

\paragraph{Verification of an optimization.} As Alive optimizations are
polymorphic over types, the Alive interpreter checks the correctness of
the optimization for each feasible type (up to a bound on integer
width). Alive models various kinds of undefined behavior in LLVM (\ie,
poison values, undef values, and true undefined
behavior)~\cite{lopes:alive:pldi:2015}. The subtleties of the
semantics are currently being explored~\cite{LLVM-undef-proposal}. For
simplicity, we use a definedness constraint for an instruction to
exclude all kinds of undefined behavior while describing the
verification below.

For each feasible type assignment, \alive creates two expressions for each
instruction: $\val$, the value it returns, and $\welldef$, the
necessary conditions for it to have well-defined behavior.
The interpreter also generates an SMT expression $\pre$ corresponding to the
precondition. A transformation is correct if and only if the
target is defined and the roots of the source and target DAGs produce the same
value when the precondition is satisfied and the source
is defined. That is:
\[
\forall_{\vin}: \pre \land \srcdef \implies \tgtdef \land
\srcval = \tgtval,
\]
where $\vin$ is the set of input variables in the DAG, $\srcdef$
and $\tgtdef$ are constraints for the source and the
target to be have defined behavior, respectively, and $\srcval$ and $\tgtval$
are the values computed by the source and target.

%% file: precondition.tex
\begin{figure}[t]
\begin{small}

  \[
  \begin{array}{l@{\,}rcl}

& \ottnt{pre} & \bnfdef & \ottnt{pred} \bnfalt \lnot \ottnt{pre} \bnfalt 
	\ottnt{pre}\, \land\, \ottnt{pre} \bnfalt \ottnt{pre} \lor \ottnt{pre} \\

& \ottnt{pred} & \bnfdef & \ottnt{binpred} \bnfalt \ottnt{pfun} \\

& \ottnt{binpred} & \bnfdef & \ottnt{cexpr}\, \ottnt{cond}\, \ottnt{cexpr}  \\

& \ottnt{cexpr} & \bnfdef & \ottmv{constant} \bnfalt \ottnt{unop}\, \ottnt{cexpr} \bnfalt \\
& & & \ottnt{cexpr}\, \ottnt{binop}\, \ottnt{cexpr} \bnfalt \ottnt{cfun} \\

& \ottnt{cond} & \bnfdef & \ottkw{eq}\ \bnfalt\ \ottkw{ne}\ \bnfalt\ \ottkw{ugt}\ \bnfalt
         \ottkw{uge}\ \bnfalt\ \ottkw{ult}\ \bnfalt \\
        & & & \ottkw{ule}\ \bnfalt
         \ottkw{sgt} \bnfalt  \ottkw{sge}\ \bnfalt \ottkw{slt}\ \bnfalt
         \ottkw{sle}\\

%& \ottnt{typ} & \bnfdef &   \ottkw{i} \ottmv{sz} 
%   \bnfalt \ottnt{typ}  \ottsym{*}
%   \bnfalt  \ottsym{[}\,  \ottmv{sz} \, \times \,  \ottnt{typ}  \, \ottsym{]} 
%  \\
& \ottnt{binop} & \bnfdef & \ottkw{add}\bnfalt\ottkw{sub}\bnfalt\ottkw{mul}\bnfalt
         \ottkw{udiv} \bnfalt\ottkw{sdiv}\bnfalt \\

& & & \ottkw{urem}\bnfalt \ottkw{srem}\bnfalt\ottkw{shl}\bnfalt\ottkw{lshr}\bnfalt
         \ottkw{ashr}\bnfalt \\
& & & \ottkw{and}\bnfalt\ottkw{or}\bnfalt
         \ottkw{xor}\\

& \ottnt{unop} & \bnfdef & \ottkw{neg} \bnfalt \ottkw{not} \\

& \ottnt{cfun} & \bnfdef & \ottkw{abs}\, \ottnt{cexpr} \bnfalt \ottkw{log2}\, \ottnt{cexpr} \bnfalt \ottkw{width}\, \ottmv{value} \\

%& \ottnt{conv} & \bnfdef &  \ottkw{zext} \bnfalt  \ottkw{sext}\bnfalt \ottkw{trunc}  \\

& \ottnt{pfun} & \bnfdef & \ottkw{isSignBit}\, \ottnt{cexpr} \bnfalt \ottkw{isPowerOf2}\, \ottnt{cexpr} \bnfalt \\
& & & \ottkw{isPowerOf2OrZero}\, \ottnt{cexpr}

  \end{array}
  \]
\vspace{4pt}
\hrule width \hsize height .33pt
\vspace{4pt}
\caption{Abstract syntax of preconditions. }
\label{fig:syntax}
\end{small}
\end{figure}

%% file: sec.approach.tex
\section{Precondition Inference}
\label{sec:inference}
\infer is an end-to-end solution that infers preconditions for
LLVM optimizations expressed in Alive.
Our approach is inspired by PIE~\cite{padhi:pie:pldi:2016}, a
data-driven approach for inferring preconditions and loop invariants
for general-purpose programs. However, we design new techniques and
algorithms to address the following challenges in the Alive context:
generating data in a static setting, handling type
polymorphism, addressing compile-time undefined behavior in
Alive, and generating succinct partial preconditions.

\infer consists of three components: (1) the example generator, which
produces a set of positive and negative examples, (2) the predicate
learner, which learns a set of predicates that can
separate the positive and negative examples, and (3) the Boolean
formula learner, which learns a Boolean formula in conjunctive normal
form (CNF).

\input{alg.outerloop} Figure~\ref{alg.outerloop} provides a high level
sketch of our approach. \infer's example generator addresses the
challenge of generating data in a static setting where input programs
are not available. It ensures that the feasible types of an
optimization are sufficiently represented in the set of examples. It
also ensures that examples do not cause any compile-time undefined
behavior. The developer can also guide the example generation process.
\infer's predicate learner synthesizes new predicates
while accounting for types in Alive and learns a set of
predicates that are useful in separating the positive and the negative
examples. It must account for safety of the
predicate at compile time. \infer's Boolean formula learner generates
either a partial or weakest precondition using the learned predicates.
The Boolean learning algorithms need to account for the
safety of the learned preconditions. \infer checks the validity of the
optimization with both partial and weakest preconditions.
If the optimization is not valid with the learned precondition, \infer
adds counterexamples to the set of negative examples and repeats the
process. When the optimization is valid but there are positive
examples that are disallowed by the precondition, \infer reports the
partial precondition to the developer and adds the positive examples to
the set of positive examples and repeats the process to weaken the
precondition.  Next, we describe compile-time undefined behavior,
which influences all components of \infer.

\subsection{Addressing Compile-time Undefined Behavior}
\label{sec:compile-time-ub}
Alive's constant expression language includes operations that are not
defined for all possible inputs. The semantics for Alive's integer
expressions are based on the corresponding semantics for SMT bitvector
operations and LLVM's arbitrary-precision integer library. Both
include operations that are undefined in certain circumstances, such as
division by zero. Verifying an optimization which is not fully defined
may result in unexpected or inconsistent solver behavior. Performing
such an optimization in LLVM using Alive-generated code may result in
a compiler crash.

For example, the precondition \texttt{C1 \% C2 == 0} is undefined
when \texttt{C2} is zero. We say it has \emph{compile-time undefined behavior},
to distinguish it from the undefined behavior present in LLVM~IR.
In particular, Alive-generated code in LLVM may crash when evaluating
the precondition.

We associate a \emph{safety condition} with each Alive term.
This condition is true if and only if the term does not have compile-time
undefined behavior. For our example, the safety condition
is $\mathtt{C2} \ne 0$. Calculation of safety conditions is mostly straightforward,
but there are some subtleties. The precondition
\texttt{C2 != 0 \&\& C1 \% C2 == 0} is always safe, because the corresponding
SMT expression is well-defined and the C++ translation will avoid dividing
by zero due to short-circuit evaluation.

We extend the refinement condition for Alive to include safety conditions.
The source of an optimization contains no constant expressions, so it is
trivially safe. The precondition must always be safe to evaluate. Any
constant expressions in the target must be safe when the precondition is
satisfied.
The new refinement condition for checking the correctness of an Alive
optimization is:
\begin{equation}
\forall_{\vin}\ \presafe \land (\pre \implies \tgtsafe \land
(\srcdef \implies \tgtdef \land \srcval = \tgtval)),
\label{eq:refinement}%
\end{equation}%
where $\pre$, $\srcval$, $\tgtval$, $\srcdef$, $\tgtdef$, $\tgtsafe$,
and $\presafe$ are constraints to represent the precondition, the
value produced by source, value produced by the target, definedness
constraints for the source, definedness conditions for the target,
safety conditions for the compile time constant expressions in the
target, and safety conditions for the constant expressions in the
precondition, respectively.  Here, $\runtimevars$ and
$\compiletimevars$ represent the set of input variables and symbolic
constants, respectively. We have changed \alivenj to use the new
refinement condition for verification. Each component of \infer has to
consider safety while learning a precondition.

\subsection{Example Generation}
\label{sec.example-generation}

One contribution of \infer is the use of a data-driven approach to
infer preconditions in the context of compiler verification, where
concrete input programs are not available. \infer has to generate
examples in this setting. The key challenges in example generation
are: handling type polymorphism in Alive, identifying a method to
classify an example as positive or negative, and methods to quickly
generate sufficient and diverse examples.

\paragraph{What is an example?}
An example in our setting represents an input program that matches the
source of an optimization.  The precondition determines when an
optimization can be performed using the information available while
compiling a concrete input program: the types of the source values and
the values of the symbolic constants. Hence, examples in \infer
contain type assignments and values for the symbolic
constants (consistent with their assigned types).

\infer uses examples with different type assignments to avoid creating
preconditions that are only valid for one assignment. For example,
some operations may overflow at small types, but not at large types.
Additionally, having examples at different types increases the chances
of learning predicates that vary based on types, such as
\texttt{width(\%a)}.

\paragraph{Positive and negative examples.}
We classify an example as \emph{positive} if the optimization's target
refines the source and has no compile-time undefined behavior when the
symbolic constants in the optimization are substituted with concrete
values from the example.  Otherwise, it is \emph{negative}. Any
example that causes compile-time undefined behavior or fails the
refinement check is a negative example, which should be disallowed by
the learned precondition.
Given sets of runtime variables $\runtimevars$ and symbolic constants
$\compiletimevars$, we simplify the refinement check in
Section~\ref{sec:compile-time-ub} and define
\begin{equation}
V(\compiletimevars, \runtimevars) \equiv \tgtsafe \land 
(\srcdef \implies \tgtdef \land \srcval = \tgtval).
\end{equation}\label{eq:ref-func}%
We use $V(e,\runtimevars)$ to represent the substitution of symbolic
constants with concrete values from the example~$e$ in the above
equation. Hence, an example $e$ is positive if and only if
$\forall_\runtimevars V(e, \runtimevars)$.

\paragraph{Methods to generate examples.}
To handle Alive's type polymorphism, we first sample the set of type
assignments for the variables in the optimization. For a given type
assignment, \infer obtains examples using three methods: using an SMT
solver, classifying randomly-generated examples, and classifying a 
small set of examples using boundary values.
We can obtain negative examples by passing the following negated refinement
condition to the SMT solver and extracting values for the symbolic
constants from the models it returns:
\begin{equation}
\exists_e \exists_\runtimevars \lnot V(e, \runtimevars).
\end{equation}%
Using an SMT solver to find positive examples is similar, but we
additionally require positive examples to have a well-defined source
for at least one assignment of run-time variables. This is not
necessary for correctness, but allowing the precondition to reject
trivial positive examples can result in simpler preconditions.
Thus, we obtain positive examples by using values for the symbolic
constants from the models of the following formula:
\begin{equation}
\exists_e
(\exists_\runtimevars \srcdef) \land (\forall_\runtimevars V(e,
\runtimevars)).
\end{equation}%

The final two methods involve first generating examples and then
classifying them. We create examples by
randomly choosing values for each symbolic constant in
$\compiletimevars$ that fall within its type. We create additional
examples by taking the Cartesian product of
$\{0,1,-1,m\}$ for each variable in $\compiletimevars$ ($m$ is the
minimum signed value for that type).
Once we obtain a proposed example $e$, we use an SMT solver to check
whether $\exists_\runtimevars
\subst{\srcdef}{\compiletimevars}{e}$, where we specialize $\srcdef$
by substituting the variables from $\compiletimevars$ with their values from $e$.
If not, then $e$ is trivial and
gets discarded. If so, we check whether $\exists_\runtimevars \lnot
V(e, \runtimevars)$. If so, $e$ is negative. Otherwise, $e$ is
positive.

The example generation methods have complementary benefits.  Random
generation of values for symbolic constants is fast but may not
generate enough positive and/or negative examples. In contrast,
examples generated using SMT solvers can be slow.

We cannot use a fixed number of examples for every optimization
because number of type assignments are exponential in the number of
feasible types for an optimization. We increase the number of examples
logarithmically with the number of
possible type assignments.

\paragraph{Developer support to guide example generation.}
\infer will eventually find a precondition which accepts all positive
examples and rejects all negative examples, but sometimes the
developer may wish to exclude examples from consideration.  Examples
may be excluded because the structure in LLVM ensures that the
optimization will never be applied to them, so the precondition need
not explicitly reject them. Conversely, some positive examples may
represent uninteresting cases, and the precondition need
not explicitly accept them. The developer can inform \infer about such
examples using assumptions, which are terms in the precondition
language. \infer discards any example that does not satisfy the
assumptions.

\input{alg.prefromexamples.tex}
\input{fig.illustration.tex}

\subsection{On-demand Predicate Enumeration and Learning}
\label{sec.on-demand-learning}

Inspired by PIE~\cite{padhi:pie:pldi:2016}, \infer separates
predicate learning from Boolean formula learning. In contrast to PIE,
\infer addresses the following challenges: handling predicates that
are unsafe with respect to an example and enumerating type-polymorphic
predicates consistent with an optimization's type constraints.
Given a set of examples, \infer creates a sample of examples that are
not currently separated by the learned predicates.
It enumerates predicates on-demand until it finds a predicate that
separates the positive and negative examples in the sample, meaning
it accepts all the positives and rejects all the negatives, or vice
versa. When it finds such a
predicate, it tests the predicate on the entire set of examples.  When
it has accumulated a set of predicates that accept some (or all)
positive examples and reject all negative examples from the set of
examples, it learns a Boolean formula for the
precondition. Figure~\ref{alg.learnfromexamples} provides a sketch of
our algorithm for predicate learning.

\paragraph{Constructing the predicate matrix.} To identify whether the
algorithm has learned a sufficient number of predicates to accept all
positive examples and reject all negative examples, \infer
conceptually constructs a \emph{predicate matrix}.
The rows in this matrix correspond to the examples and the
columns correspond to the currently learned predicates.
The matrix is updated whenever a new predicate is learned.
Figures~\ref{fig:illustration}(e) and \ref{fig:illustration}(f)
illustrate predicate matrices with one and two predicates,
respectively. Each entry in the matrix is accept ($\top$), reject
($\bot$), or unsafe ($\star$).  These correspond to the results of
evaluating the predicate after substituting the type parameters and
symbolic constants in the example: it may evaluate to true, evaluate
to false, or have compile-time undefined behavior, respectively.

In Figure~\ref{fig:illustration}(e), the predicate \texttt{C1 u<
  C2} accepts the negative example \texttt{(0, 1)} and rejects the
negative example \texttt{(4,~2)}.  Here, \texttt{u<} is the unsigned
comparison operation. In Figure~\ref{fig:illustration}(f), the
predicate \texttt{C2 /u C1 == 0} is unsafe with the example
\texttt{(0, 1)} because it causes compile-time undefined behavior
(division by zero).

\paragraph{Predicate vectors.}
Each row in the predicate matrix contains the results of applying the
learned predicates to a particular example. We call such a list of
results a \emph{predicate vector}.
The same vector may be associated with multiple
examples. We call a vector that is only associated with positive
examples a positive vector. Similarly, vectors that are only
associated with negative examples are negative vectors. Vectors that are
associated with both positive and negative examples are \emph{mixed
  vectors}.
An initial predicate matrix before any predicates have been learned
associates every example with the empty vector $\langle\rangle$, which
is mixed.

\paragraph{Generating preconditions.}
If the predicate matrix contains at least one positive vector, the
algorithm can generate a partial precondition. This precondition will
reject all negative examples, and accept some positive
examples. \infer uses the partial Boolean learner described
in Section~\ref{sec.boollearner} to find a formula that accepts some
positive vectors and rejects all negative and mixed vectors. Each
positive vector is weighted by the number of associated examples.

If the predicate matrix contains no mixed vectors, the algorithm can
generate a precondition using the complete Boolean learner.
This precondition will reject all
negative examples, and accept \emph{all} positive examples.
Once a complete precondition is found, \infer
checks if it is valid and weakest. If not, it adds the new examples and
continues.

Otherwise, the predicate matrix contains at least one mixed vector, so
\infer must learn more predicates.

\paragraph{Learning predicates.}
The purpose of introducing new predicates is to reduce the number of
examples associated with mixed vectors---ideally to zero. To do this,
\infer selects a mixed vector and searches for a predicate which
separates its associated positive and negative examples, meaning that
the predicate accepts the positive examples and rejects the negative
examples, or vice versa. If the vector has many examples, it may be
difficult to find such a predicate, so \infer searches for one which
separates ``enough'' examples. This is particularly important early on,
when all examples are associated with the empty vector. Without
sampling, \infer would have to find a single predicate which expresses
the entire precondition.

Given a mixed vector, \infer selects a certain number of examples
associated with the vector. These examples are called the \emph{sample}.
The sample always contains both positive and negative examples.
If the mixed vector has only a few examples, then all are included in
the sample.

Once the sample is selected, \infer enumerates predicates until it
finds one which separates the sample. We permit the predicate to be
unsafe for negative examples in the sample, but forbid the predicate
to be unsafe for any positive example. This simplifies precondition
generation later on, as \infer is free to assume that unsafe examples
can be rejected. If the predicate meets these conditions, it is
added to the list of learned predicates and the predicate matrix
is extended by evaluating the predicate on all examples. At this point,
\infer checks to see whether it can generate new preconditions, as
described earlier.

\paragraph{Type-aware predicate enumeration.}
\infer generates predicates using bounded recursion.
Each predicate is assigned a weight, and \infer enumerate predicates with
increasing weights.
The weight of a predicate roughly corresponds to the number of leaf
nodes in the abstract syntax tree~(AST) of the predicate in Alive's
internal representation. Exceptions are predicate functions and
constant functions, which contribute to the weight but are not leaf
nodes.

The enumerator is aware of the type constraints that should be satisfied by predicates
and expressions. For example, the arguments to a comparison must have
the same type. Enumeration of constant expressions is type-directed:
the enumerator takes the desired type expression as a parameter, it
propagates this type into subexpressions where necessary, and it
selects appropriately-typed symbols when it reaches a leaf node.
To avoid generating equivalent expressions (\eg, $a + (b+c)$,
$(a+b)+c$, $(b+a)+c$), our enumerator is aware of the algebraic
properties of the predicate language, and produces expressions in a
normal form.
However, we have to be careful in applying only algebraic identities
with bitvector arithmetic.
For example, $-(a \div b)$, $-a \div b$, and $a \div -b$ are all
distinct expressions with bitvector arithmetic.

\subsection{Boolean Formula Learning with Weighted Vectors}
\label{sec.boollearner}
\input{alg.fullboollearner.tex}
Once \infer has found a set of predicates and their behavior for each
example, it assembles these predicates into a Boolean formula using
conjunction, disjunction, and negation. \infer uses two different
methods for learning Boolean formulae. Both learn formulae that
reject all negative examples. One learns a possibly-large formula that
accepts all positive examples. The other learns a formula that
covers as many positive examples as it can while remaining succinct.
Both produce formulae in conjunctive normal form~(CNF).

The learners do not need to know about the specific examples used or
predicates learned during inference.
Instead, their inputs are the unique predicate vectors from
the predicate learner's predicate matrix. Recall that a predicate vector
is an ordered list describing the behavior of each predicate
when evaluated on an example. Figures~\ref{fig:illustration}(e--f) show
some examples of predicate vectors.
Given a set of $n$-entry predicate vectors, the learners create a formula
using abstract predicates $p_1,\ldots,p_n$, which will later be replaced
by the corresponding learned predicates. 
The behavior of $p_i$ for a vector $v$ is determined by $v_i$.
The learned formulae will accept
one or more vectors associated only with positive examples, and reject
all vectors associated with negative examples.

The simplest method for finding a Boolean formula would be to translate
each positive vector into a conjunctive clause that accepts only that vector,
and then take the disjunction of such clauses for all positive predicate
vectors.
For example, with
positive vectors $\top\bot\top$ and $\top\bot\bot$ we might learn
$(p_1 \land \lnot p_2 \land p_3) \lor (p_1 \land \lnot p_2 \land \lnot p_3)$,
which accepts only those vectors. While very simple, this method may
produce needlessly complex formulae, cannot exclude unnecessary predicates,
and does not produce formulae in CNF.

\paragraph{Evaluating clauses.}
Both learners work with clauses that are disjunctions of (possibly negated)
predicates. The function $\Call{accepts}{c,v}$ determines whether a clause
$c$ accepts or rejects a vector $v$. It evaluates $c$ by checking $v_i$ for
each $p_i \in c$, and rejects if all $p_i$ reject.

In contrast to PIE, \infer can reason about predicates that are unsafe
($\star$), meaning they exhibit undefined behavior during evaluation instead
of evaluating to accept ($\top$) or reject ($\bot$). The predicate enumerator
ensures that no predicate is unsafe for any positive vector, so the Boolean
learners are free to assume that any clause which exhibits unsafe behavior for
a vector will reject that vector. In particular, $p_i$ and $\lnot p_i$ both
reject a vector $v$ where $v_i = \star$.

Because unsafe predicates are handled by $\Call{accepts}{}$ and in the
predicate learner, the Boolean learners need not be aware of unsafe
predicates.

\input{alg.cover.tex}
\input{alg.boollearner.tex}
\paragraph{Complete Boolean formula learning.}
\infer finds formulae in two stages. First, it chooses disjunctive clauses of
up to $k$ predicates that accept all positive vectors. As shown in
Figure~\ref{alg.learnfullbooleanformula}, it begins with $k = 1$ and
iteratively increases $k$ until every negative vector is rejected by
at least one chosen clause. That is, no negative vector is accepted by
the conjunction $\bigwedge C$ of all chosen clauses in $C$.
Figures~\ref{fig.boollearner}(d--e) illustrate the process of increasing
$k$ until all negative vectors are rejected, and give the learned formula.

\input{fig.boollearner.tex}

In the next stage, \infer finds a subset of $C$ that still rejects all
negative vectors. We use a greedy approximate set-cover algorithm, shown in
Figure~\ref{alg.cover}, which repeatedly selects the clause in $C$ that
rejects the most negative vectors that have not already been rejected until
all negative vectors have been rejected.

\paragraph{Weighted partial Boolean formula learning.}
While it is always possible to find a complete Boolean formula that accepts
all positive examples and rejects all negative examples, such a formula may
be very complex. This is not always desirable, so \infer optionally
reports a set of less-complex partial preconditions, which reject all
negative examples but accept only some positive examples.

Our algorithm for finding partial Boolean formulas, shown in
Figure~\ref{alg.learnbooleanformula}, operates similarly to the complete
Boolean learner, but limits complexity by only generating disjunctive clauses of
up to $K$ predicates, where $K$ is a parameter. It first creates a set $D$ of
all clauses up to size $K$, and chooses a set $C \subseteq D$ containing
clauses that accept all positive vectors. If $\bigwedge C$ is insufficient
to reject all negative vectors, it chooses new clauses in $D$ to add to $C$.
Any new clause will reject some positive vectors. To guide the choice, \infer
associates a weight with each positive vector, and greedily chooses
clauses to maximize the total weight of the positive vectors accepted by
$\bigwedge C$. For each unselected clause, it calculates the total weight of
the positive vectors accepted by the clause, and then chooses a clause $c$ with
the highest total. Any positive vectors rejected by $c$ are discarded,
and the weight totals for the unchosen clauses are recalculated. This
continues until $\bigwedge C$ is sufficient to reject all negative clauses.
Figures~\ref{fig.boollearner}(a--c) illustrates the learner
choosing clauses and discarding positive vectors until it finds a set $C$
that rejects all negative vectors.
\infer then uses the approximate set-cover algorithm from Figure~\ref{alg.cover} to
find a subset of $C$ that rejects all negative vectors.

In our prototype, \infer uses $K = 1$ and weights predicate vectors according
to the number of associated positive examples.
We plan to investigate the impact of these heuristics as future work.
To increase the chances of finding an optimal formula, \infer may perform
this algorithm several times, making different choices for the initial selected
clause, and choosing the formula that accepts the most total weight.

%% file: alg.outerloop.tex
\begin{figure}
\begin{small}
\begin{algorithmic}
\Function{InferPrecondition}{$opt, I$}

\State $\langle E^+,E^- \rangle \gets \Call{MakeExamples}{opt}$
\State $P_{valid} \gets \emptyset$
\Repeat
  \State $\langle P_{p}, P_{f} \rangle \gets \Call{PreconditionsByExamples}{E^+,E^-,I}$
  \State $e^- \gets \emptyset$
  \ForAll{$p \in P_p$}
    \State $e_p^- \gets \Call{CounterExamples}{p, opt}$
    \If{$e_p^- = \emptyset$}
      \State $P_{valid} \gets P_{valid} \cup \{p\}$
    \EndIf
    \State $e^- \gets e^- \cup e_p^-$
  \EndFor
  \State $e_f^- \gets \Call{CounterExamples}{P_f, opt}$
  \State $e^- \gets e^- \cup e_f^-$
  \State $e^+ \gets \emptyset$
 \If{$e_f^- = \emptyset$}
   \State $P_{valid} \gets P_{valid} \cup P_f$
   \State $e^+ \gets \Call{PositiveExamples}{P_f, opt}$
 \EndIf
 \State $E^+ \gets E^+ \cup e^+$
 \State $E^- \gets E^- \cup e^-$ 
\Until{$e^- = e^+ = \emptyset$}
\State \Return $P_{valid}$

\EndFunction

\end{algorithmic}
\end{small}
\vspace{4pt}
\hrule width \hsize height .33pt
\vspace{4pt}
\caption{\small Algorithm for generating preconditions for an LLVM peephole
  optimization $opt$ with an initial set of predicates $I$. We
  generate an initial set of examples with the function
  \protect\Call{MakeExamples}{}. The function \protect\Call{PreconditionsByExamples}{} enumerates
  predicates on-demand and returns a tuple: (a set of partial
  preconditions and a complete precondition for the sample).
  Both the
  partial preconditions and the complete precondition are checked for
  validity and counter examples are added to the set of bad
  examples. If the complete precondition is valid, it checks if it is
  the weakest.}
\label{alg.outerloop}
\end{figure}

%% file: alg.prefromexamples.tex
\begin{figure}
\begin{small}
\begin{algorithmic}

\Function{PreconditionsByExamples}{$E^+, E^-, I$}

\State $P \gets I$
\State $M \gets \Call{EmptyPredicateMatrix}{}$
\ForAll{$p \in I$}
  \State $M \gets \Call{AddPredicate}{p, M}$
\EndFor

\State $\Phi \gets \emptyset$

\While{$\Call{MixedVectors}{M} \ne \emptyset$}
  
  \State $V_w^+, V^- \gets \Call{WeightedPartition}{M}$
  \State $\phi \gets \Call{LearnPartialBoolean}{P, V_w^+, V^-, 1}$
  \State $\Phi \gets \Phi \cup \{\phi\}$
  
  \Statex
  
  \State Select $v \in \Call{MixedVectors}{M}$
  \State $e^+, e^- \gets \Call{Sample}{v, M}$
  \State $p \gets \Call{LearnPredicate}{e^+, e^-}$
  \State $P \gets P \cup \{p\}$
  \State $M \gets \Call{AddPredicate}{p, M}$
\EndWhile

\State $V^+, V^- \gets \Call{Partition}{M}$
\State $\phi_f \gets \Call{LearnCompleteBoolean}{P, V^+, V^-}$
\State \Return $\langle \Phi, \phi_f \rangle$
\EndFunction
\end{algorithmic}
\end{small}
\vspace{4pt}
\hrule width \hsize height .33pt
\vspace{4pt}
\caption{\small Algorithm for learning preconditions given a set of
  examples and an intial set of
  predicates~($I$). Function~\protect\Call{AddPredicate}{} adds a
  predicate to the predicate
  matrix. Function~\protect\Call{WeightedPartition}{} partitions the
  predicate vectors into positive vectors and negative vectors and the
  weight of the positive vector is the number of positive examples
  accepted by the positive
  vector. Function~\protect\Call{LearnPartialBoolean}{} computes the
  partial precondition using the weighted positive vectors and
  negative vectors~(see Figure~\ref{alg.learnbooleanformula}). When
  the predicate matrix does not have any mixed vectors, the weakest
  precondition is computed by the
  function~\protect\Call{LearnCompleteBoolean}{}~(see
  Figure~\ref{alg.learnfullbooleanformula}). The algorithm returns a tuple
  --- a set of valid partial preconditions and the weakest
  precondition --- for the given set of examples.}
\label{alg.learnfromexamples}
\end{figure}

%% file: fig.illustration.tex
\begin{figure*}
  \centerline{\includegraphics[height=3in]{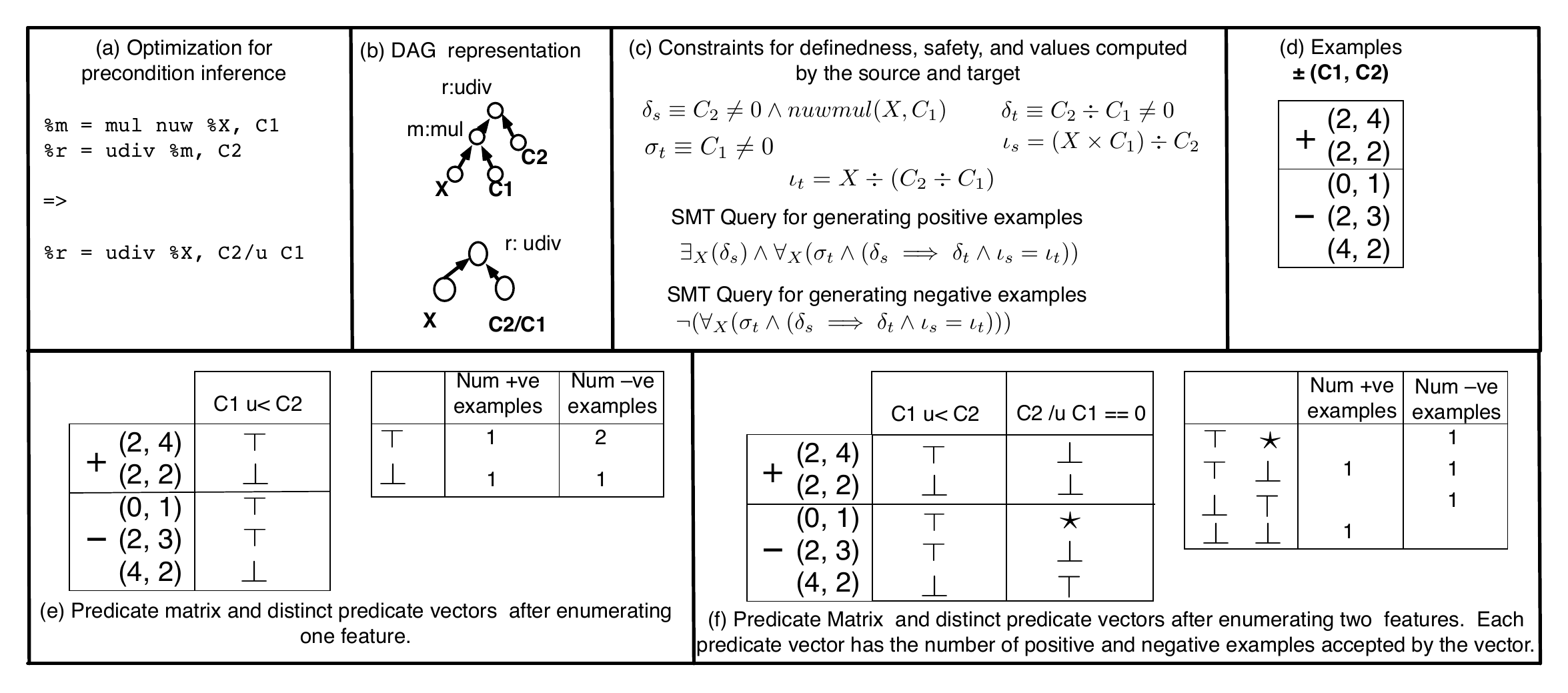}}
  \hrule width \hsize height .33pt
  \vspace{4pt}

\caption{ \small The process of learning preconditions. (a) LLVM
  peephole optimization expressed in Alive whose precondition is being
  learned. (b) DAG representation of the optimization, which has input
  runtime variable $X$ and symbolic constants \texttt{C1} and
  \texttt{C2}. (c) Constraints for the definedness of the
  source~($\srcdef$), definedness of the target~($\tgtdef$), compile-time
  safety of the target~($\tgtsafe$), value produced by the
  source~($\srcval$), and value produced by the
  target~($\tgtval$). Queries provided to SMT solvers to generate
  positive and negative examples are also provided. The predicate
  $nuwmul(X, C_{1})$ encodes the fact that $X$ multiplied by
  $C_1$ can be represented as an unsigned integer in the current type
  assignment.
  (d)~Sample set of examples generated. We omit type assignments for
   simplicity. An example
  \texttt{(4, 2)} represents a positive example with $C_1 = 4$
  and $C_2 = 2$. Any example with $C_2 = 0$ will be
  discarded, as it causes undefined behavior in the source.  In
  contrast, any example with $C_1 = 0, C_2 \ne 0$ will be
  marked negative, because it causes an unsafe computation in the
  target. (e) The predicate matrix and distinct predicate vectors
  after adding the predicate \texttt{C1~u<~C2}. $\top$ indicates
  that the predicate accepts the example. $\bot$ indicates that the
  predicate rejects the example. $\star$ indicate that the example is
  unsafe (compile-time undefined behavior). (f) Predicate matrix after
  adding two features. Our incomplete boolean learner will find a
  subset of the predicates \texttt{C1 u< C2}, \texttt{C1 u>=
    C2}, \texttt{C2 /u C1 == 0}, and \texttt{C1 /u C2~!=~0},
  which accepts as many of the positive vectors as it can and rejects
  all the negative and mixed vectors.  For this matrix, it will
  produce the precondtion \texttt{(C1 u>= C2) \&\& (C2 /u C1 != 0)}.
  }
\label{fig:illustration}
\end{figure*}

%% file: alg.fullboollearner.tex
\begin{figure}
\begin{small}
\begin{algorithmic}
\Function{LearnCompleteBoolean}{$Preds, V^+, V^-$}
\State $lits \gets Preds \cup \{ \lnot p : p \in Preds \}$
\State $k \gets 0$
\State $C \gets \emptyset$
\While{$\exists v \in V^- \text{ s.t. } \accepts{\bigwedge C}{v}$}
  \State $k \gets k + 1$
  \State $C_k \gets \{ \bigvee d : d \subseteq lits, |d| = k \}$
  \State $C \gets C \cup \{ d : d \in C_k,  \forall v \in V^+ (\accepts{d}{v}) \}$
\EndWhile
\State \Return $\Call{CoverClauses}{C, V^-}$
\EndFunction
\end{algorithmic}
\end{small}
\vspace{4pt}
\hrule width \hsize height .33pt
\vspace{4pt}
\caption{\small Algorithm for learning a Boolean formula given a set of
predicates, positive vectors, and negative vectors.}
\label{alg.learnfullbooleanformula}
\end{figure}

%% file: alg.cover.tex
\begin{figure}
\begin{small}
\begin{algorithmic}
\Function{CoverClauses}{$C, V^{-}$}
\State $P \gets \top$
\While{$\exists v \in V^- \text{ s.t. } \accepts{P}{v}$}
  \State $c \gets \argmax_{d \in C} |\{v : v \in V^-, \lnot\accepts{d}{v}\}|$
  \State $V^- \gets V^- \setminus \{v : v \in V^-, \lnot\accepts{c}{v}\}$
  \State $C \gets C \setminus \{c\}$
  \State $P \gets P \land c$
\EndWhile
\State \Return $P$
\EndFunction
\end{algorithmic}
\end{small}
\vspace{4pt}
\hrule width \hsize height .33pt
\vspace{4pt}
\caption{\small Greedy set-cover algorithm that returns a set of
  clauses rejecting all negative examples.}
\label{alg.cover}
\end{figure}

%% file: alg.boollearner.tex
\begin{figure}
\begin{small}
  \begin{algorithmic}
    \Function{LearnPartialBoolean}{$Preds$, $V_w^+$, $V^-$, $K$}
  \State $lits \gets Preds \cup \{ \lnot p : p \in Preds \}$
  \State $D \gets \{ \bigvee d : d \subseteq lits, |d| \le K \}$

  \State $C \gets \emptyset$ 
  \While{$\exists v \in V^- \text{ s.t. } \accepts{\bigwedge C}{v}$}
    \State $A \gets \{ \langle w,v \rangle : \langle w,v \rangle \in V_w^+, \accepts{\bigwedge C}{v} \}$
    \State $c \gets \argmax_{d \in D} \sum \{ w : \langle w,v \rangle \in A, \accepts{d}{v} \}$
    \State $C \gets C \cup \{c\}$
    \State $D \gets D \setminus \{c\}$
    \If{$D = \emptyset$}
    	\State \Return $\bot$
    \EndIf
  \EndWhile
  \State \Return $\Call{CoverClauses}{C, V^-}$
  
\EndFunction
\end{algorithmic}
\end{small}
\vspace{4pt}
\hrule width \hsize height .33pt
\vspace{4pt}
\caption{\small Algorithm for learning a partial Boolean formula that rejects all
  negative vectors and maximizes the weights of the positive vectors
  accepted. }
\label{alg.learnbooleanformula}
\end{figure}

%% file: fig.boollearner.tex
\begin{figure*}

\scalebox{0.75}{%
\begin{minipage}{0.44\textwidth}
\begin{tabular}{lr|cccccc}
$V^+$ & $w$ & $p_1$ & $\lnot p_1$ & $p_2$ & $\lnot p_2$ & $p_3$ & $\lnot p_3$ \\
\hline
$\bot\top\bot$ &  8 &   & + & + &   &   & + \\
$\bot\top\top$ &  8 &   & + & + &   & + &   \\
$\top\bot\bot$ & 10 & + &   &   & + &   & + \\
$\top\bot\top$ &  1 & + &   &   & + & + &   \\
$\top\top\top$ &  3 & + &   & + &   & + &   \\
\hline
               &    &14 &16 &19 &11 &12 &18 \\
$V^-$ &&&&&&& \\
\hline
$\bot\bot\bot$ &    &$-$&   &$-$&   &$-$&   \\
$\bot\bot\top$ &    &$-$&   &$-$&   &   &$-$\\
$\top\top\bot$ &    &   &$-$&   &$-$&$-$&   \\
\end{tabular}

(a) Select $p_2$, discard $\top\bot\bot$, $\top\bot\top$, $\bot\bot\bot$ and $\bot\bot\top$.
\end{minipage}
}
\scalebox{0.75}{%
\begin{minipage}{0.44\textwidth}
\begin{tabular}{lr|cccccc}
$V^+$ & $w$ & $p_1$ & $\lnot p_1$ & $p_2$ & $\lnot p_2$ & $p_3$ & $\lnot p_3$ \\
\hline
$\bot\top\bot$ &  8 &   & + & + &   &   & + \\
$\bot\top\top$ &  8 &   & + & + &   & + &   \\
               &    &   &   &   &   &   &   \\
               &    &   &   &   &   &   &   \\
$\top\top\top$ &  3 & + &   & + &   & + &   \\
\hline
               &    & 3 &16 &   & 0 &11 & 8 \\
$V^-$ &&&&&&& \\
\hline
               &    &   &   &   &   &   &   \\
               &    &   &   &   &   &   &   \\
$\top\top\bot$ &    &   &$-$&   &$-$&$-$&   \\
\end{tabular}

(b) Select $\lnot p_1$, discard $\top\top\top$ and $\top\top\bot$.
\end{minipage}
}
\scalebox{0.75}{%
\begin{minipage}{0.44\textwidth}
\begin{tabular}{lr|cccccc}
$V^+$ & $w$ & $p_1$ & $\lnot p_1$ & $p_2$ & $\lnot p_2$ & $p_3$ & $\lnot p_3$ \\
\hline
$\bot\top\bot$ &  8 &   & + & + &   &   & + \\
$\bot\top\top$ &  8 &   & + & + &   & + &   \\
               &    &   &   &   &   &   &   \\
               &    &   &   &   &   &   &   \\
               &    &   &   &   &   &   &   \\
\hline
               &    & 0 &   &   & 0 & 8 & 8 \\
$V^-$ &&&&&&& \\
\hline
               &    &   &   &   &   &   &   \\
               &    &   &   &   &   &   &   \\
               &    &   &   &   &   &   &   \\
\end{tabular}

(c) Final: $p_2 \land \lnot p_1$.
\end{minipage}
}

\vspace{1em}

\scalebox{0.75}{%
\begin{minipage}{0.3\textwidth}
\begin{tabular}{l|c}
$V^-$ & $p_1 \lor p_2$ \\
\hline
$\bot\bot\bot$ & $-$ \\
$\bot\bot\top$ & $-$ \\
$\top\top\bot$ &  \\
\end{tabular}

(d) 2-CNF terms
\end{minipage}
}
\scalebox{0.75}{%
\begin{minipage}{0.6\textwidth}
\begin{tabular}{l|cccc}
$V^-$ & $p_1 \lor p_2$ & $p_1 \lor p_2 \lor p_3$ & $p_1 \lor p_2 \lor\lnot p_3$ & $\lnot p_1 \lor\lnot p_2 \lor p_3$\\
\hline
$\bot\bot\bot$ & $-$ & $-$ &     &    \\
$\bot\bot\top$ & $-$ &     & $-$ &    \\
$\top\top\bot$ &     &     &     & $-$\\
\end{tabular}

(e) 3-CNF terms. Cover is $(p_1 \lor p_2) \land (\lnot p_1 \lor\lnot p_2 \lor p_3)$.
\end{minipage}
}
\vspace{4pt}
\hrule width \hsize height .33pt
\vspace{4pt}
\caption{\small Illustration of the partial and complete Boolean
  learners on the same predicate matrix. In each table, rows
  correspond to vectors and columns correspond to clauses. A $+$
  indicates that the clause accepts a positive vector, and a $-$
  indicates that it rejects a negative vector. (a--c) show how the
  partial learner selects clauses until all negative clauses are
  rejected.  In (a), the algorithm selects $p_2$ as it maximizes the
  weight and it discards positive vectors $\top\bot\bot$,
  $\top\bot\top$, $\bot\bot\bot$, and $\bot\bot\top$ because $p_2$
  rejects them. (d--e) show how the complete learner adds larger
  clauses until all negative clauses are rejected. Any clause
  considered by the complete learner has to accept all positive
  vectors. Only clauses which accept all positive vectors are shown in
  this figure.  }
\label{fig.boollearner}
\end{figure*}

%% file: sec.generalize.tex
\section{Generalizing Concrete Expression DAGs}
\label{sec:generalize}
\input{fig.generalization.tex}
To further demonstrate the applicability of \infer, we generalize
optimization patterns generated by Souper~\cite{souper,souper-blog}, an
LLVM~IR--based superoptimizer.
 The initial prototype of
Souper collects a database of expression DAGs that evaluate to
either true or false~\cite{souper-blog}. It also generates concrete
path conditions with such expression DAGs. We focus on expression DAGs
without path conditions, because they can be translated to Alive.  New
peephole optimizations have been added to LLVM based on the patterns
discovered by Souper~\cite{souper-llvm}. In such scenarios, developers
typically prefer to add a generalized version.

We create a generalized version of a Souper DAG by replacing all concrete
constants in the source of the optimization with symbolic
constants. However, we cannot replace a concrete constant in the
target with a symbolic constant because Alive does not allow the
definition of new symbolic constants in the target.
Figure~\ref{fig:generalization}(1a) and
Figure~\ref{fig:generalization}(2a) present the expression DAGs in
Alive syntax.
The generalized optimization and preconditions generated by \infer are
shown in Figure~\ref{fig:generalization}(1b) and
Figure~\ref{fig:generalization}(2b).
To illustrate, the weakest precondition generated by \infer for the
generalized optimization in Figure~\ref{fig:generalization}(1b) is
\begin{small}
\begin{verbatim}
(C3 != 0 || C2 == 0) && 
(C2 u<= 1 || (C4 & ~C1) != 0 || C4 < 0) && 
((C4 & ~C1) != 0 || C4 >= C2) && 
(C3 != 0 || C1 == 0) && 
(C2 != 0 || C4 == 0 || (C4 & ~C1) !=0) && 
(C2 u> 1 || C4 u<= 1 || (C4 & ~C1) != 0) && 
(isSignBit(C4) || C2 + 1 >= 0 || (C4 & ~C1) != 0)
\end{verbatim}
\end{small}
Unfortunately, this weakest precondition is not succinct. \infer
also generated a partial precondition,
\begin{verbatim}
C4 & ~C1 != 0 && C3 != 0
\end{verbatim}
also shown in Figure~\ref{fig:generalization}(1b). It is succinct and accepts 95\%
of the positive examples, which makes a case for generating partial
preconditions.

%% file: fig.generalization.tex
\begin{figure*}[ht!]
  \centerline{\includegraphics[width=\linewidth]{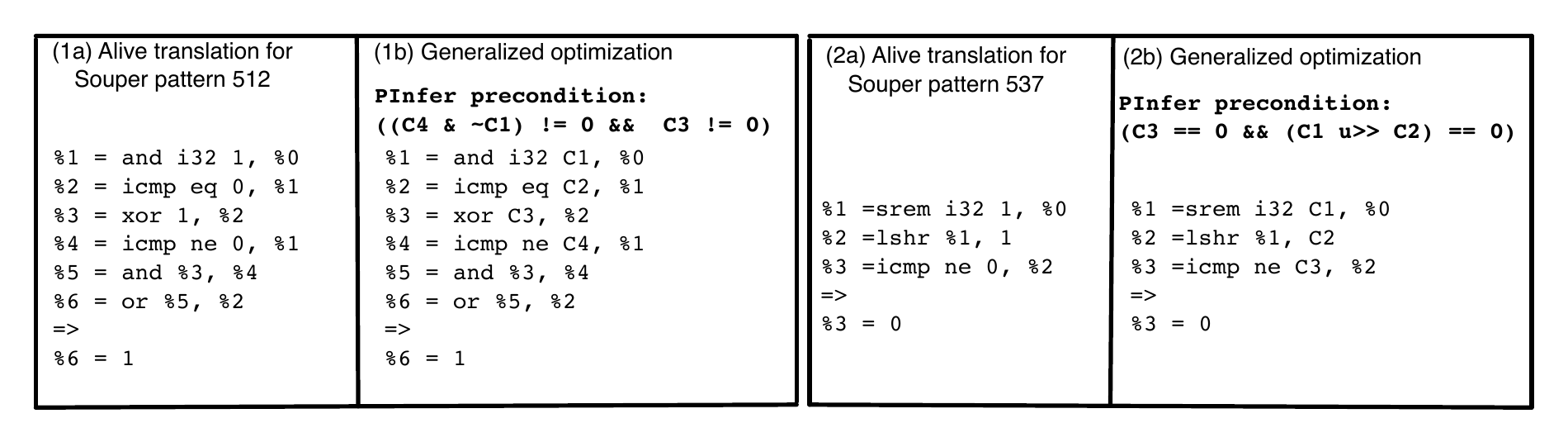}}
  \hrule width \hsize height .33pt
    \vspace{4pt}
\caption{\small Generalization of optimization patterns generated by
  Souper with \infer. (a)~Alive version of the Souper
  pattern. (b)~Generalized optimization with all concrete values in
  the source replaced by symbolic constants along with the inferred
  precondition.  }
\label{fig:generalization}
\end{figure*}

%% file: sec.evaluation.tex
\section{Evaluation}
\label{sec.evaluation}
We describe the \infer prototype, our methodology, and our experience
inferring preconditions for LLVM peephole optimizations. Our
experiments evaluate the effectiveness of the \infer prototype in
generating both weakest and partial preconditions.

\paragraph{The \infer prototype.}
We built the \infer prototype by extending the publicly available
\alivenj toolkit~\cite{alivenj-git}.
\alivenj also supports verification of floating point optimizations,
but we leave precondition inference for those to future work.

\infer enhances \alivenj with three major features. (1)
Implementations of the enumeration and learning
algorithms, comprising roughly two thousand lines of Python code. (2)
Safety analysis, which expresses the conditions under which an
optimization target or precondition may have undefined behavior at
compile-time. (3) Separation of the type checking and type
assignment phases. \infer assigns each term an abstract type once
during type checking or predicate enumeration. These abstract types
are then mapped to concrete types during validation without the need
of re-performing type checking.

In our experiments, we use Z3~4.4.1~\cite{z3solver} to handle SMT
queries. The \infer prototype is open source and publicly available as
part of \alivenj toolkit.

\paragraph{Optimization suite.}
We use 417 optimizations from the \alive suite, a snapshot of
optimizations from LLVM's InstCombine and InstructionSimplify passes.
Some preconditions in Alive are weaker than LLVM's preconditions.
Of these 417 optimizations, 195 require no precondition and 41 rely on
dataflow analyses for runtime values.  For the 195 optimizations that
do not have a precondition, \infer successfully infers
\texttt{true}. \infer does not support 41 optimizations that use
dataflow analyses for runtime values.
Of the remaining 181 that have a precondition in the \alive suite,
seven require constant functions or predicates not currently supported
by \infer.
This leaves us with 174 optimizations for which \infer could possibly
derive preconditions.

%

%\vspace{4pt}
\paragraph{Methodology.} 
In our experiments for precondition inference, we removed the
precondition in the optimization and provided it to the \infer
prototype. We compare the precondition generated by the \infer
prototype and the original precondition for it in Alive (to determine
if it is weaker or stronger).  All experiments were performed on a
64-bit Intel Skylake--processor machine with four cores and 16~GB of
RAM.

\paragraph{Precondition search}
To perform a fair assessment of the benefits of precondition inference
using on-demand predicate learning, we
created a variation of \infer that enumerates all possible preconditions
until it finds a valid, weakest precondition. We refer to this method
as precondition search, and call our modified prototype Alive-Search.

Alive-Search generates preconditions in CNF, using the predicate
enumerator as a subroutine. A precondition's size is the sum of
the sizes of its predicates. All preconditions of a given size are
generated before any precondition of the next larger size.

Each precondition is tested against a set of examples
generated using the method from Section~\ref{sec.example-generation}.
If the precondition accepts all positive instances and does not accept
any negative instances, it is then verified by the SMT solver.
If the solver finds counter examples, or additional positive examples
which the precondition rejects, testing continues with the next
precondition.
It is never necessary to reconsider a previously-rejected precondition.

This algorithm has two advantages over precondition inference: it always
finds a precondition of minimum size, and it can never get stuck with
a difficult-to-separate sample.
%\footnote{A bad set, in this case, would be one
%  which is divided too easily, leading to extra work for the solver.}
%
The disadvantage is that this algorithm cannot break the search
problem into smaller parts.
The number of preconditions and predicates for a given size both grow
exponentially, with the number of preconditions growing somewhat
faster.
Consider an optimization for which the minimally-sized precondition
has two predicates of sizes $m$ and $n$.
The number of preconditions which must be searched will be
$O(c^{m+n})$, which is vastly larger than $O(c^m + c^n)$, the
best-case amount of work needed for the predicate learner to find the
two predicates.
The exponential growth means that the predicate learner is still
faster even if it wastes most of its effort learning predicates which
will later be discarded.

\input{fig.predicates}
\subsection{Effectiveness in Generating Preconditions}
We test the effectiveness of our approach by generating preconditions
for the optimizations in the Alive suite. We ran inference for each
optimization with a 1000-second timeout.

\infer successfully generated weakest
preconditions for 133 out of the 174 optimizations.  Although \infer
could not generate weakest preconditions for the remaining 41
optimizations, it generated partial preconditions for 31
optimizations.
For six optimizations, it was not able to learn sufficient predicates
to generate any preconditions within the timeout period.
In one case, the Boolean learner failed to find a formula.
Finally, in three cases Z3 returned unknown during example generation or in
final validation.
In summary, \infer was able to
generate either the weakest or a partial precondition for 164 out of
the 174 optimizations.

Figure~\ref{fig:predicates} provides summary information on the number
of predicates in the generated weakest precondition, number of
distinct predicates in the weakest precondition, the total number of predicates
learned, and maximum disjunction size in the final formula
learned by the Boolean learner. Figure~\ref{fig:predicates}(a) shows
that about 80 optimizations in the suite
have a single predicate in the precondition, and around 40 optimizations
have two to four predicates. Figure~\ref{fig:predicates}(b) shows
that the number of distinct predicates in the weakest precondition is
lower than the number of predicates, which is common in formulae
expressed in conjunctive normal form.

Figure~\ref{fig:predicates}(c) characterizes the number of predicates
needed to separate the examples for various optimizations.
\infer learned no more than 28 predicates for any optimization.
Often, not all learned predicates were needed in the final formula.
Learned predicates may occur multiple times in a precondition,
so the maximum precondition size is greater than the largest
number of learned predicates.
Figure~\ref{fig:predicates}(d) shows that only 45 optimizations
required disjunction to express a weakest precondition, indicated
by having a clause size greater than one.

The number of predicates enumerated over the course of predicate learning
varies widely. For nine optimizations where \infer found a weakest
precondition, the learner considered more than
ten thousand predicates and as many as 104,000. For nineteen, it
considered between one and ten thousand. For sixteen,
between a hundred and a thousand. For eighty-six, between eleven and a hundred.
For four, between one and ten.

\paragraph{Preconditions with generalization.}
To evaluate the applicability of \infer with generalization, we
translated 71 concrete optimization instances from the initial results
of Souper to Alive and generalized them with symbolic
constants. \infer was able to generate weakest
preconditions for 51 optimizations. It generated partial preconditions
for additional 3 optimizations. In the remaining 17 cases,
Z3 hung while generating positive
examples, Z3 returned unknown, or \infer could not learn a Boolean
formula that rejects all negative vectors.

\input{fig.infertime}
\subsection{Comparison of \infer with Alive-Search}
For comparison, we also experimented with precondition search
(described earlier in this section) for same set of optimizations. In
contrast to \infer, Alive-Search was
able to generate weakest preconditions for 114 of 174
optimizations. It performs similarly to \infer for
optimizations with few predicates in the precondition. It times out
for optimizations that have more than three predicates in the
precondition.

Figure~\ref{fig.infertime} reports the number of optimizations for
which \infer and Alive-Search
were able to generate weakest preconditions within a given amount of
time.
Both tools
generate weakest preconditions in 10 seconds (per optimization) for
about 100 optimizations.
These optimizations have one or two predicates in the
precondition~(see Figure~\ref{fig:predicates}(a)).
In summary, we observe that predicate inference can find
preconditions for more optimizations than predicate search
within a given time limit.

\input{fig.weak.tex}
\subsection{Strength of Inferred Preconditions}
We compare weakest preconditions generated by \infer with the
preconditions in the Alive suite. We consider a precondition
weakest if it accepts every example
where the optimization is non-trivially valid; i.e., the source is
well-defined for some run-time input.  To determine whether a
precondition generated by \infer ($\phi_{I}$) accepts more examples than
its counterpart in the suite ($\phi_{A}$), we check the satisfiability
of the formula $\phi_{I} \land \lnot \phi_{A}$ using an SMT solver.

Of the 133 optimizations where \infer is able to generate a
weakest precondition, 73 are weaker than the suite's
precondition. Figure~\ref{fig:weak} shows four of these optimizations,
with the preconditions from the suite and generated by \infer.
For the remaining 61 optimizations,
our prototype generated a precondition equivalent to its suite
counterpart. Even the partial
preconditions generated were often weaker than the suite preconditions.
Of the partial preconditions generated, 15 are
incomparable because there are examples which are accepted by the
\infer precondition but not by the suite precondition and vice-versa.
These include the example in Section~\ref{sec:intro}.

The structure of LLVM's peephole optimization pass creates implicit
assumptions for many optimizations: if two optimizations can apply
to the same input program, only the first will be applied.
\infer
learns preconditions in isolation; assumptions must be explicit.
Even taking these implicit assumptions into consideration, we have
found optimizations where \infer finds a weaker precondition
and reported them to LLVM's developers~\cite{infer-llvm-dev}.
We plan to investigate others in the
future.

%% file: fig.predicates.tex
\begin{figure*}
  \centerline{\includegraphics[width=\textwidth]{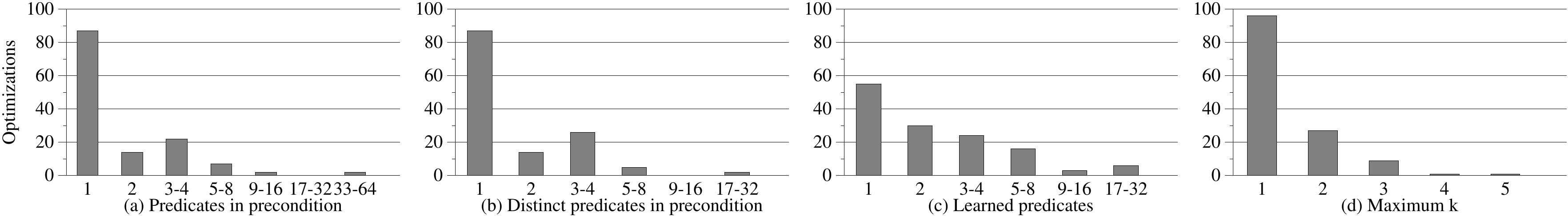}}
\vspace{4pt}
\hrule width \hsize height .33pt
\vspace{4pt}
\caption{\small Information about weakest preconditions
  successfully inferred within 1000~s. The histograms show the number
  of optimizations with (a) the number of predicates in the
  precondition, (b) the number of distinct predicates in the
  precondition (a predicate and its negation are not considered
  distinct), (c) the number of predicates accepted by the learner
  during inference, and (d) the maximum number of predicates occuring
  in a disjunction (\ie, the value of $k$ reached by the Boolean
  formula learner).  }
\label{fig:predicates}
\end{figure*}

%% file: fig.infertime.tex
\begin{figure}
\centerline{\includegraphics{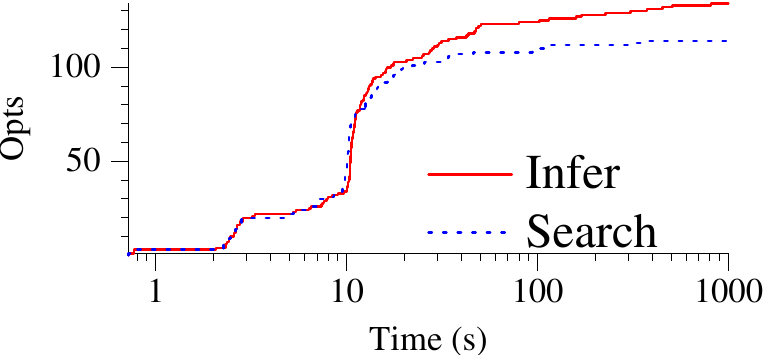}}
\vspace{4pt}
\hrule width \hsize height .33pt
\vspace{4pt}
\caption{\small Number of optimizations for which a weakest
  precondition was inferred by precondition inference
  (Infer) and precondition search (Search) within a given time
  limit. The $x$-axis is running time, in seconds, used to infer the
  precondition of each optimization. The $y$-axis is the cumulative
  number of optimizations which required at most that time.
  }
\label{fig.infertime}
\end{figure}

%% file: fig.weak.tex
\begin{figure*}
  \centerline{\includegraphics[width=\linewidth]{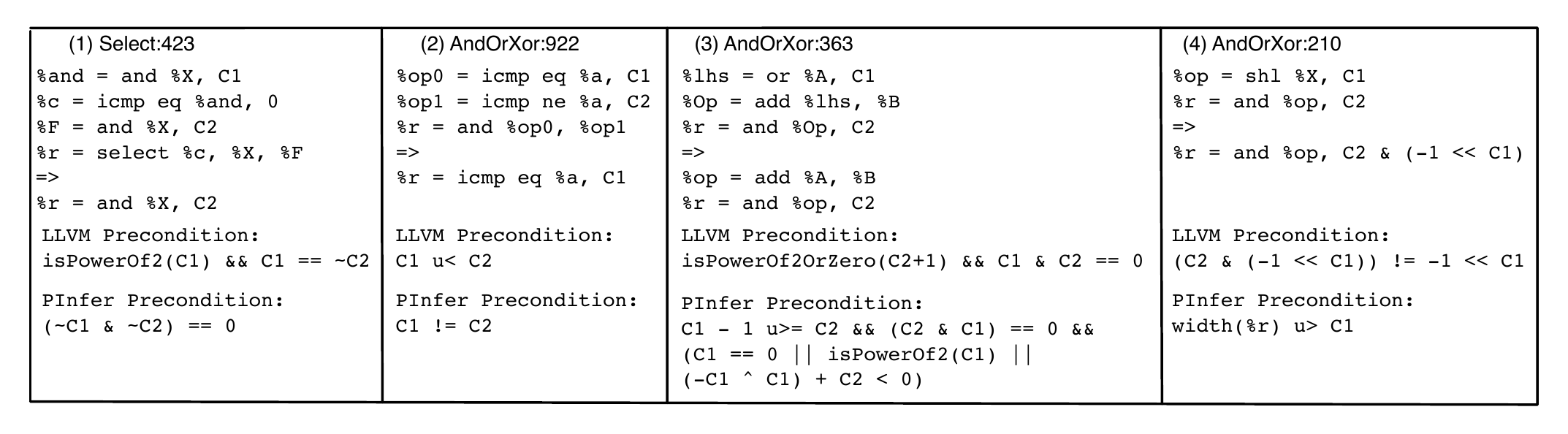}}
%\vspace{4pt}
\hrule width \hsize height .33pt
\vspace{4pt}
\caption{\small A sample of optimizations where \infer generated a
  weaker precondition compared to the precondition in LLVM/Alive. We
  provide the name of the optimization in the Alive suite, the
  LLVM/Alive precondition and the \infer precondition. (1) Consider
  the instance \texttt{C1 = 3, C2 = 14} for 4-bit integers, \ie,
  \texttt{0011} and \texttt{1110}. These satisfy neither of the
  clauses in LLVM's precondition, but do satisfy the \infer's
  precondition, which can be rewritten as \texttt{C1 | C2 == -1}. (2)
  This optimization's source calculates $a = C_1 \land a \ne C_2$ and
  the target $a = C_1$. By the transitive property, this is equivalent
  to $a = C_1 \land C_1 \ne C_2$. \infer generates the equivalent
  precondition \texttt{C1 > C2 || C1 < C2}. (3) Consider the instance
  \texttt{C1 = 10, C2 = 2} for 4-bit integers, \ie, \texttt{1100} and
  \texttt{0010}. This is rejected by LLVM's precondition, because
  three is not a power of two, but is accepted by \infer's. (4)
  \infer's precondition is clearly weaker, as it will accept the cases
  where \texttt{C2} is masked by \texttt{-1 << C1} as long as
  \texttt{C1} is less than the bit width. For example, \texttt{C1 = 2,
    C2 = 14} for 4-bit integers, \ie, \texttt{0001} and \texttt{1110}.
 }
\label{fig:weak}
\end{figure*}

%% file: sec.related.tex
\section{Related Work}
There is a large body of work on inferring specifications ---
preconditions, postconditions, and invariants --- for general purpose
programs~\cite{daikon:2007,padhi:pie:pldi:2016,sankaranarayan:dt:issta:2008,
  martin:commutative-spec:cav:2015, garg:ice:cav:2014,
  garg:ice-dt:popl:2016,seghir:cegar:esop:2013, dillig:oopsla:2013,
  interface:popl:2005, mulitiabduction:popl:2016,
  mining-specifications:2002,weakest-unstructured-programs:paste:2005,
  sharma:guess-and-check:esop:2013,cousot:vmcai:2013}. Data-driven
approaches have also been explored for inferring
specifications~\cite{padhi:pie:pldi:2016,sankaranarayan:dt:issta:2008,
  martin:commutative-spec:cav:2015,garg:ice:cav:2014,
  garg:ice-dt:popl:2016}.
We primarily focus on closely related work in this section.

\paragraph{PIE.} Our work is inspired by
PIE~\cite{padhi:pie:pldi:2016}, which generates preconditions for
general-purpose programs. PIE uses predicate learning, which it calls
feature learning, along with a Boolean learner to separate positive
and negative examples. \infer differs from PIE by addressing
challenges specific to LLVM and Alive. First, we identify the need for
succinct partial preconditions and propose a weighted partial Boolean
formula learner. Second, we propose a strategy to generate positive
and negative examples while handling polymorphic types and
compile-time undefined behavior. Third, we design a predicate learner
which can reason about predicates with potential compile-time
undefined behavior.

\paragraph{Compiler precondition synthesis.} Prior approaches have also
explored precondition generation for compiler
optimizations~\cite{scherpelz:rhodium:pldi:2007,lopes:vmcai:2014,Buchwald:cc15}.
PSyCO~\cite{lopes:vmcai:2014} synthesizes read-write preconditions
given a finite predicate set.  They do not address the complexities of
bitvector arithmetic and the interaction with undefined behavior.
Optgen~\cite{Buchwald:cc15} automatically generates all peephole
optimizations within a specified size bound and verifies their
correctness. These optimizations may include preconditions, which are
expressions of the form \texttt{expr == 0} and are found using enumeration.

\paragraph{Logical abduction methods.} Another approach to
precondition inference is logical
abduction~\cite{giacobazzi:ilps:1994, dillig:oopsla:2013}. Methods
using quantifier elimination \cite{dillig:oopsla:2013} are promising,
but methods for eliminating quantifiers in bitvector algebra work only
for a small subset of operations~\cite{john:qe-lmc:icms:2014}.  We
initially tried logical abduction methods by restricting optimizations
to use only linear integer arithmetic (LIA) but settled on a
data-driven approach to increase its applicability.

\paragraph{Data-driven inference methods.}
Other prior data-driven approaches often work only with predefined
predicates~\cite{sankaranarayan:dt:issta:2008,
  martin:commutative-spec:cav:2015,
  sharma:guess-and-check:esop:2013}. Researchers have used
counter-example guided refinement~\cite{clarke:cegar:cav:2000},
similar to \infer, by beginning with overlapping positive and negative
sets and refining them by finding
counter-examples~\cite{seghir:cegar:esop:2013}. They also require a
fixed set of predefined predicates. ICE and ICE-DT
\cite{garg:ice:cav:2014, garg:ice-dt:popl:2016} use positive,
negative, and implication examples for synthesizing invariants. They
use either a template-based synthesis or a decision tree learning
algorithm to generate invariants using a fixed set of attributes.
\infer, similar to PIE, learns and synthesizes predicates on-demand.
\paragraph{Search techniques and superoptimization.}  \infer's inference
can be viewed as a variant of various symbolic, stochastic, and
enumerative search strategies employed in program
synthesis~\cite{solar:asplos:2006, gulwani:pldi:2011,
  torlak:pldi:2014, komuravelli:cav:2016, alur:fmcad:2013} and 
  superoptimizers~\cite{phothilimthana:greenthumb:asplos:2016,Bansal06,schkufza13,
  Massalin87,joshi06}. \infer can be used to generalize/validate
patterns generated by superoptimizers.

\paragraph{Compiler correctness.} A compiler can be written in a
mathematical theorem prover (\eg,
CompCert~\cite{leroy:compcert:jar:2009},
Vellvm~\cite{zhao:vminus:pldi:2013,zhao:vellvm:popl:2012}), which
would require one to figure out the specification in such a
setting~\cite{mullen:peek:pldi:2016,
  tatlock:xcert:pldi:2010}. Alternatively, various other DSLs have
also been proposed for compiler
construction~\cite{kundu:pec:pldi:2009, lerner:rhodium:popl:2005}.
\infer generates preconditions or optimizations expressed in
Alive~\cite{lopes:alive:pldi:2015}.  In principle, \infer can apply to
other DSLs.  Our recent work has explored compiler non-termination
errors with a suite of peephole optimizations~\cite{alive-loops},
which typically occurs when profitability metrics are not included in
the precondition. The weakest preconditions inferred by \infer should
be checked with those tools before including them in LLVM to avoid
non-termination errors.

%% file: sec.conclusion.tex
\section{Conclusion}
We show that it is possible to infer preconditions for peephole
optimizations in LLVM using a data-driven approach with on-demand predicate
learning.
We highlight the trade-off between applicability and succinctness of
the precondition.
The \infer prototype addresses the challenges of polymorphic types and
compile-time undefined behavior in the precondition language to
generate both weakest and succinct partial preconditions.
Our goal is to assist LLVM developers in debugging an invalid
optimization.
\infer is likely to be useful to LLVM developers, as it is able to
generate preconditions weaker than LLVM's preconditions.